\documentclass[aip,jap,10pt,twocolumn]{revtex4-1}

\usepackage{amsmath}
\usepackage{graphicx}
\usepackage{epstopdf}

\newcommand{\pdiff}[2]{\frac{\partial #1}{\partial #2}}

\begin{document}

\title{Sensitive Magnetic Force Detection with a Carbon Nanotube Resonator}
\date{\today}
\author{Kyle Willick}
\affiliation{Institute for Quantum Computing, University of Waterloo, Waterloo, Ontario, Canada}
\affiliation{Waterloo Institute for Nanotechnology, University of Waterloo, Waterloo, Ontario, Canada}
\affiliation{Department of Physics and Astronomy, University of Waterloo, Waterloo, Ontario, Canada}
\author{Chris Haapamaki}
\affiliation{Institute for Quantum Computing, University of Waterloo, Waterloo, Ontario, Canada}
\affiliation{Department of Chemistry, University of Waterloo, Waterloo, Ontario, Canada}
\author{Jonathan Baugh}
\email[Contact: ]{baugh@iqc.ca}
\affiliation{Institute for Quantum Computing, University of Waterloo, Waterloo, Ontario, Canada}
\affiliation{Waterloo Institute for Nanotechnology, University of Waterloo, Waterloo, Ontario, Canada}
\affiliation{Department of Physics and Astronomy, University of Waterloo, Waterloo, Ontario, Canada}
\affiliation{Department of Chemistry, University of Waterloo, Waterloo, Ontario, Canada}

\begin{abstract}
We propose a technique for sensitive magnetic point force detection using a suspended carbon nanotube (CNT) mechanical resonator combined with a magnetic field gradient generated by a ferromagnetic gate electrode. Numerical calculations of the mechanical resonance frequency show that single Bohr magneton changes in the magnetic state of an individual magnetic molecule grafted to the CNT can translate to detectable frequency shifts, on the order of a few kHz. The dependences of the resonator response to device parameters such as length, tension, CNT diameter, and gate voltage are explored and optimal operating conditions are identified. A signal-to-noise analysis shows that in principle, magnetic switching at the level of a single Bohr magneton can be read out in a single shot on timescales as short as 10~$\mu\text{s}$. This force sensor should enable new studies of spin dynamics in isolated single molecule magnets, free from the crystalline or ensemble settings typically studied.  

\end{abstract}

\maketitle

\section{Introduction}
The ability to sense magnetic moments on the order of the Bohr magneton would enable novel studies of magnetism at the single molecule level. While a few recent experiments have achieved sensitivity near these levels at 1 Hz bandwidths \cite{Ganzhorn2013a, Vasyukov2013}, these techniques are not sensitive enough to permit high bandwidth, high fidelity readout of single molecule magnetic states. Moreover, high field studies of nanomagnetic objects are often of interest, and are not accessible in nano-SQUID experiments \cite{Vasyukov2013}, for example. Here we aim to develop sufficiently fast sensing at the Bohr magneton level to enable single-shot magnetic state readout in the presence of a magnetic field of arbitrary strength. Since high sensitivity and fast system response are competing properties, optimal tradeoffs must be achieved through careful design and analysis of the sensor. 

Suspended carbon nanotube (CNT) resonators have been used in a wide range of high sensitivity measurements \cite{Lassagne2008,Chaste2012,Moser2013,Meerwaldt2012}, where they excel due to a high Young's modulus and extremely low mass in comparison to other nanoelectromechanical systems. In a suspended CNT field-effect transistor geometry, a bottom gate electrode is used to both drive and sense resonant motion. The mechanical oscillations of the CNT modulate the capacitance to the gate, and are translated to conductance fluctuations that can be observed in mixing \cite{Sazonova2004,Gouttenoire2010} or direct-current measurements \cite{Huttel2010}. Recent experiments have demonstrated resonance frequency measurements at the intrinsic resonator noise limit at 1.2~K \cite{Moser2013}, and submicrosecond readout times at room temperature using a low-noise HEMT amplifier \cite{Meerwaldt2013}. Hence, CNT resonators can allow both high sensitivity and wide bandwidth, making them attractive candidates for sensing at the single molecule level.

\begin{figure}[b]
\centering
\includegraphics{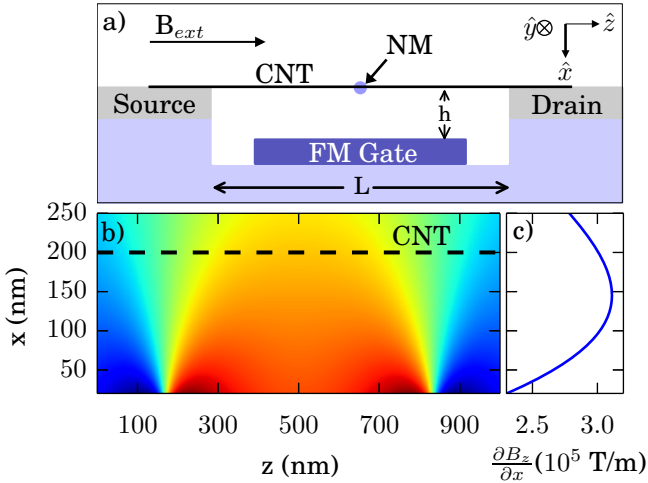}
\caption[FIXME]{\textbf{a)} Schematic device geometry, showing the ferromagnetic (FM) gate combined with a CNT electromechanical resonator. \textbf{b)} Magnetic field $B_z$ along z generated by a 200~nm thick by 800~nm wide Cobalt gate at saturation magnetization, $10^6$ A/m. \textbf{c} The field gradient $\pdiff{B_z}{x}$ at the center of the suspended CNT channel.}
\label{fig:CNTDiagram}
\end{figure}

Previous magnetic sensing with CNT resonators has exploited the torque exerted by a magnetic field on nano-objects with anisotropic magnetic moments, to shift the resonant frequency \cite{Lassagne2011,Ganzhorn2013a}. In this paper, we propose an approach using a ferromagnetic (FM) gate to generate a strong field gradient at the nanomagnet position, giving rise to a magnetic point force on the CNT. This point force generation is similar to magnetic resonance force microscopy \cite{Sidles1995,Poggio2010}, however the sensitivity of the CNT removes the requirement for magnetic resonance in the sample of interest. Figure \ref{fig:CNTDiagram}(a) shows a schematic representation of the device geometry under consideration. The FM gate generates a magnetic field gradient on the order of $10^5\, \text{T/m}$ at a height $h \sim 200$ nm above the surface of the gate, the nominal height we choose for the CNT based on fabrication requirements. Figure \ref{fig:CNTDiagram}(b) shows the component of magnetic field along the CNT axis, $B_z$, generated by a typical cobalt gate at saturation magnetization \cite{Sharma2007,Aus1998}.
Figure \ref{fig:CNTDiagram}(c) shows the field gradient $\pdiff{B_z}{x}$ at the center of the suspended CNT.

\section{Model}
The nanomagnet (NM) of interest is attached to the CNT, and experiences a force due to the magnetic field gradient
\begin{equation}
\vec{F}_{mag} = \nabla \left( \vec{m} \cdot \vec{B} \right)
\end{equation}
where $\vec{m}$ is the NM magnetic moment vector and $\vec{B}$ is the magnetic field at the NM position. Assuming the extent of the gate in the y-direction is very long compared to the device length $L$, and the NM is approximately centered above the FM gate, then $\pdiff{B_y}{x} = 0$ and $\pdiff{B_x}{x} \approx 0$. Then, the force exerted by the NM is linearly proportional to the moment along z and acts perpendicular to the CNT axis, in the vertical (x) direction: 
\begin{equation} \label{eq:Fmag}
\vec{F}_{mag} \approx m_z \pdiff{B_z}{x} \hat{x}
\end{equation}

To determine the effect of this force on the mechanical resonance, the suspended CNT is modelled as a doubly-clamped cylindrical Euler-Bernoulli beam, with a uniform electrostatic force due to the gate and a point force due to the NM. Experimental data from similar devices \cite{Witkamp2006} has shown good agreement with Euler-Bernoulli models, and the CNT-metal interface at the source and drain contacts provides sufficient force to justify the clamping assumption \cite{Cao2005}. Under this model, the governing equation for CNT motion is given by
\begin{equation} \label{eq:GoverningEquation}
\begin{split}
- E I \pdiff{^4 x}{z^4} + T \pdiff{^2 x}{z^2} + K_{elec} + F_{mag} \delta(z - z_0) - \eta \pdiff{x}{t} \\ = \rho A \pdiff{^2 x}{t^2}
\end{split}
\end{equation}
where $E$ is the Young's modulus of the CNT, $I = \frac{\pi}{4} r^4$ is the moment of inertia, $r$ is the CNT radius, $x$ is the vertical displacement of the CNT from equilibrium, $T = T_0 + \frac{E A}{2 L} \int_0^L \left(\pdiff{x}{z}\right)^2 dz$ is the tension in the CNT, $T_0$ is the residual tension at zero applied force as a result of fabrication, $K_{elec} = \frac{1}{2} C'_g V_g^2$ is the electrostatic force per unit length, $C'_g = \frac{2 \pi \epsilon_0}{h (\ln(2 h / r))^2}$ is the derivative of the CNT-gate capacitance per unit length with respect to $x$, $z_0$ is the NM position along the CNT, $\eta$ is the damping factor per unit length, $\rho$ is the CNT mass density, and $A$ is the cross-sectional area of the CNT.

The gate voltage is decomposed into a large DC voltage and a small AC component, $V_g^{dc}$ and $V_g^{ac} \cos(\omega t)$ respectively, such that $V_g^{ac} \ll V_g^{dc}$. The resulting electrostatic force on the CNT can then be described by a large DC component and the first order AC term, $F_{elec} = F^{dc}_{elec} + F^{ac}_{elec} \cos(\omega t) \equiv \frac{1}{2} C'_g \left( \left(V_{g}^{dc}\right)^2 + 2 V_{g}^{dc}  V_{g}^{ac} \cos(\omega t)\right)$. The CNT displacement has a similar decomposition $x(z,t) = u(z) + v(z,t)$, where the steady state response will be of the form $v(z,t) = v(z)\cos(\omega t + \phi)$, with phase $\phi$ depending on resonator and driving conditions. Finally, the tension in the CNT will also have a large static component and an oscillatory component, which we approximate to first order as $T = T_{dc} + T_{ac}$,
\begin{gather}
T_{dc} = T_0 + \frac{E A}{2 L} \int_0^L \left(\pdiff{u}{z}\right)^2 dz \\
T_{ac} = \left(\frac{E A}{L} \int_0^L \left(\pdiff{u}{z}\pdiff{v}{z}\right) dz \right)
\end{gather}
Substituting these decompositions into Equation (\ref{eq:GoverningEquation}) gives governing equations for the DC and first-order AC components of the CNT motion.
\begin{gather}
\begin{split}
- E I \pdiff{^4 u}{z^4} + T_{DC} \pdiff{^2 u}{z^2} + K_{elec}^{dc} + F_{mag}\delta(z - z_0) = 0 \label{eq:DCequation}
\end{split}\\
\begin{split}
- E I \pdiff{^4 v}{z^4} + T_{DC} \pdiff{^2 v}{z^2} + T_{AC} \pdiff{^2 u}{z^2} + K_{elec}^{ac} \cos(\omega t)\\ - \eta \pdiff{v}{t} = \rho A \pdiff{^2 v}{t^2} \label{eq:ACequation}
\end{split}
\end{gather}
Equations (\ref{eq:DCequation}) and (\ref{eq:ACequation}) can be solved numerically, following an analysis similar to Ref.  \citenum{Witkamp2009}, to calculate the mechanical resonance frequency\cite{Supplemental}.

\section{Results}
The mechanical resonance frequency will shift depending on the strength of the magnetic point force, hence, readout of magnetic moment states depends on the ability to distinguish these frequency shifts. We calculate the frequency shift due to a magnetic moment reversal of a single Bohr magneton ($\Delta m = 2\mu_b$) as a benchmark. An example CNT resonator, called Device A, is considered with: $L = 1\:\mu\text{m}$, $r = 0.5\:\text{nm}$, the NM centered over the gate ($z_0 = \frac{L}{2}$), zero residual tension ($T_0 = 0$), and a gate-CNT separation of $200\:\text{nm}$, which gives a field gradient of $3\cdot 10^5\, \text{T/m}$. We find numerically that the shift in resonance frequency due to a change in the NM state is maximized for a DC gate voltage of 59~mV for Device A. The maximum frequency shift is $\Delta f_0 = 6.3\:\text{kHz}$ for $\Delta m = 2\mu_b$. Comparatively, to achieve a 4~kHz frequency shift with a torque-based magnetometer using the same CNT device parameters requires $\Delta m  = 24\mu_b$, when the contact length between the CNT and NM is $0.5\:\text{nm}$ \cite{Lassagne2011}.

\subsection{Length and Diameter Dependence}
\begin{figure}[b]
\centering
\includegraphics{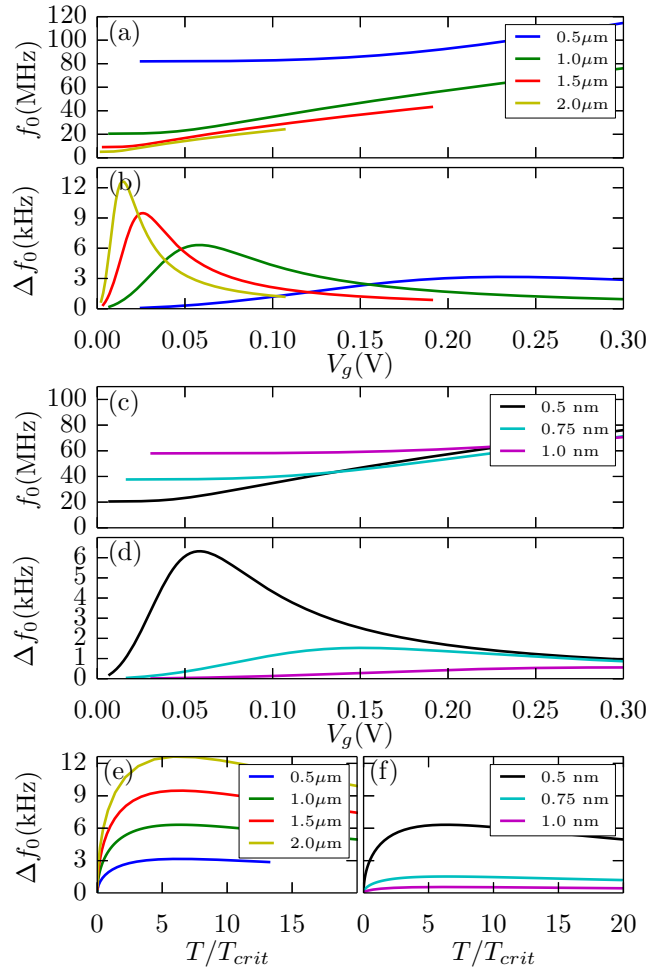}
\caption[FIXME]{\textbf{a} The resonance frequency versus gate voltage for various suspended CNT lengths, using $r=0.5\:\text{nm}$, $h=200\:\text{nm}$, $\pdiff{B_z}{x} = 3\cdot10^5\,\text{T/m}$. Note that all resonator behaviour is symmetric with respect to the sign of $V_g$. Numerical precision limits the range over which some curves are displayed. \textbf{b} The gate voltage dependence of the resonance frequency shift produced by a NM magnetic transition of $\Delta m = 2\mu_b$, for the devices from (a).
\textbf{c} The resonant frequency of CNTs with various radii, using $L=1\:\mu\text{m}$, $h=200\:\text{nm}$, $\pdiff{B_z}{x} = 3\cdot10^5\,\text{T/m}$. \textbf{d} The frequency shift of the devices in (c). 
\textbf{e},\textbf{f} The frequency shifts from (b) and (d) respectively, versus tension scaled by the critical tension, $T_{crit}$.}
\label{fig:designDependence}
\end{figure}

The resonance frequency shift is dependent on many device parameters, including length, CNT radius, and gate voltage. Figures \ref{fig:designDependence}(a) and (b) show the calculated frequencies and frequency shifts, respectively, as a function of applied gate voltage for several resonator lengths, $L$. For any fixed $L$, there is a maximum frequency shift at a small, finite 
DC gate voltage, $V^{max}_g$. As $L$ increases, $V^{max}_g$ decreases while the magnitude of the frequency shift increases linearly \cite{Supplemental} with $L$. Comparing Figures \ref{fig:designDependence}(a) and (b), we see that a larger frequency shift corresponds to a lower resonance frequency. This can be understood  as longer CNT devices having lower spring constants, yielding both lower resonance frequencies and a larger displacement in response to an applied force. 

The mechanical behaviour is very sensitive to the CNT radius, as the bending rigidity is proportional to $r^4$. Figure \ref{fig:designDependence}(d) shows the frequency shift response for a 1~$\mu\text{m}$ long CNT with three different radii. Smaller CNTs produce larger frequency shifts that occur at lower gate voltages. The maximum frequency shift ($\Delta f_0$ that occurs at $V^{max}_g$) is proportional to $r^{-7/2}$. Figure \ref{fig:designDependence}(c) shows the resonance frequency for the same devices plotted in (d). 

The value of $V^{max}_g$ can be understood by assessing the tension induced in the CNT by the gate voltage. Figures \ref{fig:designDependence}(e) and \ref{fig:designDependence}(f) show the frequency shift for the same devices as in (b) and (d), respectively, now plotted as a function of the gate-induced tension scaled by the critical tension, $T_{crit}=\frac{E I}{L^2}$. At the critical tension, the restoring force from bending rigidity and tension, corresponding to the first two terms in Equation (\ref{eq:DCequation}), are of equal magnitude. For the full range of CNT lengths and diameters considered here, the maximum frequency shift is found to occur at approximately $T \approx 6.3\,T_{crit}$. Thus, the maximum frequency shift  occurs at a specific balance of restoring forces in the CNT, independent of the length and radius. This optimal working point can be reached for a wide range of devices by appropriate tuning of the gate voltage. Furthermore, this point can be approximately identified in the resonant frequency response of the CNT resonator, as a specific point along the transition between constant resonant frequency at low gate voltages and linear gate voltage dependence at high gate voltages.

\subsection{Residual Tension}
\begin{figure}
\centering
\includegraphics{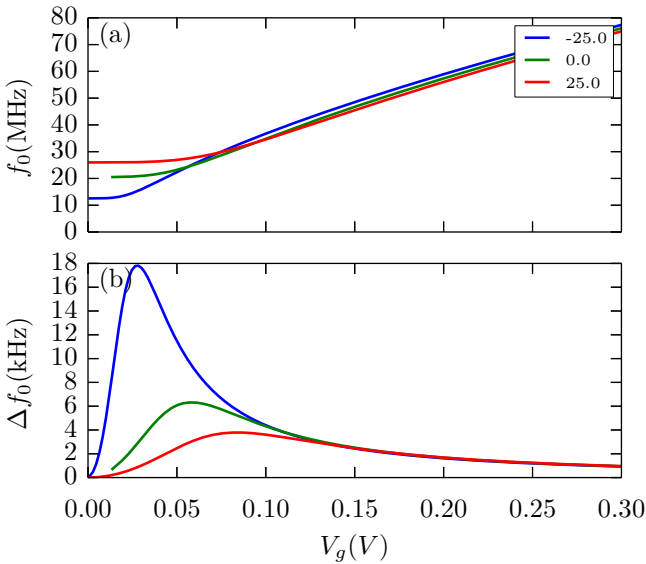}
\caption[FIXME3]{\textbf{a} Resonant frequency and \textbf{b} frequency shift using three values for residual CNT tension $T_0 / T_{crit}$, and device parameters $L = 1\:\mu\text{m}$, $d=1\:\text{nm}$, $h=200\:\text{nm}$, $\pdiff{B_z}{x} = 3\cdot10^5\,\text{T/m}$}
\label{fig:t0Effect}
\end{figure}

The fabrication and growth of suspended CNT devices can result in residual tension built into the nanotube. Until now we have considered zero residual tension, however a range of residual tensions have been observed in the experimental literature, including \cite{Witkamp2006,Huttel2009,Lassagne2009} $T'_0 \equiv T_0/T_{crit} = -26,-18,0,1$. Figure \ref{fig:t0Effect}(b) displays the frequency shift response for various residual tensions. Large residual compression, $T'_0 < -4\pi^2$, would result in buckling of the CNT \cite{Poot2009} and is not considered here. However, residual compression smaller than the buckling limit ($-4\pi^2 < T'_0 < 0$) results in a significantly increased frequency shift compared to $T'_0 = 0$. Fitting \cite{Supplemental} the maximum frequency shift as a function of $T'_0$, for various lengths and diameters, with $T'_0$ ranging from -35 to 35, we find that the maximum frequency shift is approximately proportional to $\left(4\pi^2 + T'_0\right)^{-1}$. Figure \ref{fig:t0Effect}(a) shows the resonance frequency of the devices in \ref{fig:t0Effect}(b). As in Figure 2, a larger frequency shift is again seen to correspond to a lower resonance frequency. However, at voltages above $V^{max}_g$, the effects of residual tension are reduced and the three curves join at high voltages. 

\subsection{Lateral Position Dependence}
\begin{figure}
\centering
\includegraphics{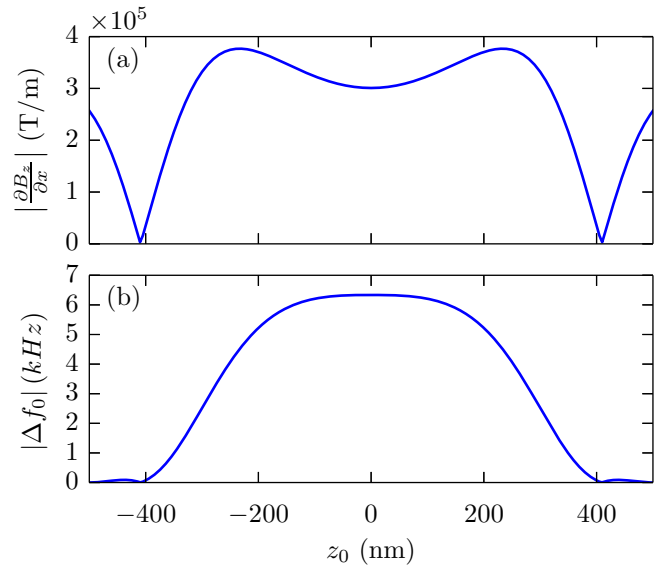}
\caption[FIXME4]{\textbf{a} The absolute magnetic field gradient $\pdiff{B_z}{x}$ as a function of position along the CNT, for the 800~nm wide by 200~nm thick cobalt gate and 200~nm gate-CNT separation. \textbf{b} The maximum frequency shift of Device A for $\Delta m = 2 \mu_B$, as a function of NM position.}
\label{fig:posDep}
\end{figure}

The position of the NM on the CNT will also influence the sensitivity to changes in the magnetic moment. We first consider the force on the NM itself. Figure \ref{fig:posDep}(a) shows the gradient of $B_z$ as a function of the position along the CNT, for Device A and the gate geometry from Figure \ref{fig:CNTDiagram}(b). As a result of this gradient, the force on the NM will increase slightly as the NM is moved away from center, before decreasing rapidly as the NM approaches the FM edge. Furthermore, the resonator response will depend on the position of the point force, via Equation  \ref{eq:GoverningEquation}.  Figure \ref{fig:posDep}(b) shows the maximum frequency shift for $\Delta m = 2 \mu_b$, as a function of position. A NM positioned anywhere in the central half of the CNT has a maximum frequency shift of at least $50\%$ of the optimal value attained for a centered NM ($z_0 = L/2$). This illustrates the robustness of this measurement technique to uncertainty in NM positioning.

\subsection{Noise and Measurement Bandwidth}
To observe the shift in resonance frequency due to a magnetic transition of the NM, the frequency shift must exceed the noise in a real frequency measurement. In usual suspended small-bandgap CNT devices using Pt contacts at temperatures near 1.2~K, the gate voltages considered here correspond to high conductance hole transport, as a result of large positive offset voltages and reduced tunnel barriers for hole transport in Pt-CNT contacts \cite{Biercuk2008}. In this hole transport regime, single electron effects such as Coulomb blockade which often dominate low-temperature suspended CNT measurements are small or absent \cite{Meerwaldt2012}. The intrinsic noise of the resonator is then dominated by thermomechanical noise \cite{Moser2013}, 
which can be modeled as a stochastic force on the resonator, with a spectral density \cite{Ekinci2004}
\begin{equation}
S_f(\omega) = 8 \pi k_b \Theta \frac{f_0}{Q}
\end{equation}
where $\Theta$ is the resonator temperature, $Q$ is the quality factor, and $f_0$ is the resonance frequency. Given recent demonstration of electrical readout of CNT motion at the thermomechanical noise limit \cite{Moser2013} at 1.2K, we assume this as the dominant noise source.

For a signal averaging time $\tau$ satisfying $\tau \gg \frac{Q}{2 \pi f_0}$, the minimum observable frequency shift is given by \cite{Ekinci2004}
\begin{equation} \label{eq:df}
\delta f_0 = \frac{1}{(2\pi)^2} \sqrt{\frac{k_B \Theta}{m' \left\langle v_{max}^2 \right\rangle \tau f_0 Q}}
\end{equation}
where $m' = m \left(\frac{1}{L} \int_0^L v^2 \: dz\right) / v_{max}^2 \approx m/2$ is the effective resonator mass, $v_{max}$ is the maximum of $v$ along $z$, and $\left\langle v_{max}^2 \right\rangle$ is the time-averaged square amplitude of the maximum AC displacement.
As the measurement averaging time is decreased, the minimum observable frequency shift approaches the infinite bandwidth limit
\begin{equation} \label{eq:df_inf}
\left.\delta f_0\right|_{\tau=0} = \frac{1}{\left(2 \pi\right)^2} \sqrt{ \frac{k_B \Theta}{m' \left\langle v_{max}^2 \right\rangle Q^2} }
\end{equation}
Using the high Q factors measured in previous experiments\cite{Huttel2009}, $Q = 10^5$, along with $\Theta = 100\:\text{mK}$, and $\left\langle v_{max}^2 \right\rangle = \left(0.1\:\text{nm}\right)^2$, the minimum detectable frequency shift signal of Device A at the optimal gate voltage ($V_g = 59\:\text{mV}$), with a 1~Hz measurement bandwidth is 5.4~Hz, as given by Equation (\ref{eq:df}). In the infinite bandwidth limit, given by Equation (\ref{eq:df_inf}), the minimum detectable frequency shift for the same device and gate setting is 86~Hz. Thus, the $\Delta m = 2\mu_b$ transition, which results in $\Delta f_0 = 6.3\:\text{kHz}$, far exceeds intrinsic noise for arbitrarily short measurement times.

The achievable measurement time for resonant frequency shifts will be limited by the mechanical response of the CNT to these frequency changes. The transient response of the mechanical resonator is characterized by a ring-down timescale, $\tau_m \sim \frac{Q}{4\pi f_0}$. For Device A, and $Q = 10^5$, the ring down time is $\tau_m \sim 300\: \mu\text{s}$. Given that cryogenic amplifiers typically operate at MHz to GHz bandwidths, the electrical measurement bandwidth should reach this limit \cite{Meerwaldt2013}, so that we expect the mechanical ring down time to be the bandwidth limiting timescale in these devices.

Table \ref{tab:deviceModels} displays the ring down times and calculated signal to noise ratios for fast readout of five example devices, where signal to noise is defined as
\begin{equation}\displaystyle
\text{SNR}_\infty = \max_{V_g} \frac{\Delta f_0}{\left.\delta f_0 \right\vert_{\tau=0}}
\end{equation}
and $\left. \delta f_0 \right|_{\tau=0}$ is evaluated using $\left\langle v_{max}^2 \right\rangle = \frac{Q V_g^{dc} V^{ac}_g C'_g}{4 \pi^2 m' f_0^2}$, and $V^{ac}_g = 1\:\mu\text{V}$.
The Devices B-E each have one parameter altered with respect to Device A. For each device, two quality factor models are considered. In one case, a fixed value of $Q = Q^*$ is used, where $Q^*=10^5$ for Devices A-D. 
The other case uses a fixed damping factor $\eta = \frac{2 \pi f_0}{Q^*}$. In both cases,  increasing length and residual compression, and decreasing diameter, improve signal to noise ratio. In the case of constant quality factor, the increased signal to noise comes at the cost of increased mechanical ring down time. Decreasing quality factor, as in Device E, decreases SNR$_\infty \propto Q^{-1}$ while linearly decreasing ring down time. The signal to noise ratio and ring down times may be adjusted through device design to obtain sufficient signal at a desired time scale.

\begin{table}[h]
\begin{tabular}{ c c  c  c  c  c }
\hline
\textbf{Device} & \textbf{Change}	& \textbf{SNR$^1_\infty$} & \textbf{SNR$^2_\infty$} & \textbf{$\tau^1_m$ ($\mu$s)} & \textbf{$\tau^2_m$ ($\mu$s)} \\ \hline
A & None & 62 & 31 & 320 & 160 \\ 
B & $L=2\mu\text{m}$ & 700 & 86 & 1300 & 160 \\ 
C & $d=1.5\text{nm}$ & 11 & 10 & 170 & 160 \\ 
D & $T'_0 = -25$ & 220 & 67 & 520 & 160 \\ 
E & $Q^*=3000$ & 1.9 & 0.9 & 9.6 & 4.8 \\ \hline
\end{tabular}
\caption{Infinite bandwidth signal to noise ratios for $\Delta m = 2\mu_b$, and ring down times, for example CNT devices. Device A has $L = 1\:\mu\text{m}, d = 1\:\text{nm}, T'_0 = 0, Q^* = 10^5$. Devices B-E have one parameter changed with respect to Device A. $\text{SNR}^1_\infty$ and $\tau^1_m$ use $Q=Q^*$. $\text{SNR}^2_\infty$ and $\tau^2_m$ use $\frac{f_0}{Q} = \frac{50\:\text{MHz}}{Q^*}$.}
\label{tab:deviceModels}
\end{table}

\section{Conclusion}
The technique presented here for magnetic point force sensing generates larger frequency shifts than previously demonstrated, predicting the highest sensitivity magnetic moment measurements to date. Additionally, the sensitivity of this technique can be tuned by altering device parameters, such as device length and residual compression, and it is robust to imprecision of the NM lateral placement along the CNT. Furthermore,
this method does not require a distribution of forces along the CNT, so it is insensitive to the grafting length between the CNT and NM, permitting a wider range of NMs to be studied. Finally, the state dependent frequency shift in this method depends only on the magnetic field gradient, and is independent of the external magnetic field once the ferromagnet is saturated. This should permit investigation of NMs at high field, which is often crucial for a full understanding of the spin physics. 

One intriguing application of this CNT-NM field gradient measurement scheme is towards the study of single molecule magnets (SMMs). The CNT-torque measurement has been previously used to examine magnetic reversal in TbPc$_2$ SMMs \cite{Ganzhorn2013a}. Beyond magnetic reversal studies, the gradient measurement scheme can be applied towards single shot readout of SMM magnetic states. Coherent spin dynamics have been demonstrated in a number of SMMs including V$_{15}$ \cite{Yang2012}, Cr$_7$Mn and Cr$_7$Ni \cite{Ardavan2007}, and Fe$_8$ \cite{Bahr2007,Bal2008,Takahashi2009}. 
Driving of the coherent transitions with an external microwave field is compatible with this technique, as long as the driving frequencies are well separated from the CNT mechanical frequencies.
The g-factors of the SMM spins are approximately $g \approx 2$, meaning that the $\Delta S = 1$ transitions observed in those experiments are approximately equivalent to the $\Delta m = 2 \mu_b$ transitions we consider. Therefore, coherent SMM spin rotations might be observable within the parameters described in this paper. SMM force measurements at the timescales described in Table \ref{tab:deviceModels} could facilitate unprecedented studies of spin dynamics at the single molecule level. 

At 1.2~K, the Cr$_7$Ni and Cr$_7$Mn SMMs have spin relaxation times ($T_1$) on the order of 1 ms in bulk crystals \cite{Ardavan2007}. If this relaxation timescale is maintained or exceeded for individual SMMs grafted to CNTs, the devices considered here might enable high fidelity spin state measurements. This single shot readout could be combined with Ramsey fringe experiments, for example, to evaluate spin decoherence and relaxation timescales of individual SMMs. These experiments could explore relaxation mechanisms by observing the dependence of the relaxation time on the applied magnetic field, and on the mechanical resonator properties such as a 1D phonon density of states. Furthermore, the ring down time scale of Device E presented in table \ref{tab:deviceModels} is comparable the $4\:\mu\text{s}$ spin decoherence time achieved with deuterated Cr$_7$Ni \cite{Ardavan2007}, potentially permitting readout on the timescale of decoherence.


\acknowledgments{We thank Jonathan Friedman for helpful discussions. This work was supported by NSERC, and the Ontario Ministry for Research and Innovation. KW gratefully acknowledges support from the Waterloo Institute for Nanotechnology.}

\clearpage

\renewcommand{\theequation}{S\arabic{equation}}
\setcounter{equation}{0}  
\renewcommand{\thefigure}{S\arabic{figure}}
\setcounter{figure}{0}
\begin{widetext}
\section*{SUPPLEMENTARY INFORMATION: Sensitive Magnetic Force Detection with a Carbon Nanotube Resonator}

\subsection*{Solving for the resonance frequency}
This solution method is based on extending the work in Ref. \citenum{Witkamp2009} to include point forces.

Starting from the Euler-Bernoulli beam model with a uniform electric force and point force from the nanomagnet
\begin{equation} \label{eqS:GoverningEquation}
- E I \pdiff{^4 x}{z^4} + T \pdiff{^2 x}{z^2} + K_{elec} + F_{mag} \delta(z - z_0) - \eta \pdiff{x}{t} = \rho A \pdiff{^2 x}{t^2}
\end{equation}

Recalling the DC and first order AC decompositions,
\begin{align}
x(z,t) = u(z) + v(z,t) \\
K_{elec} = K_{elec}^{dc} + K_{elec}^{ac} e^{i \omega t} \\
T = T_{dc} + T_{ac}(t) \\
T_{dc} = T_0 + \frac{E A}{2 L}\int_0^L \left(\pdiff{u}{z}\right)^2 dz \label{eqS:T_dc} \\
T_{ac}(t) \approx \frac{E A}{L}\int_0^L \left(\pdiff{u}{z}\pdiff{v}{z}\right) dz
\end{align}
where the term proportional to $\left(\pdiff{v}{z}\right)^2$ is neglected in $T_{ac}$.

Substituting the decompositions into Equation (\ref{eqS:GoverningEquation}) and collecting the DC and first order AC terms gives
\begin{align}
- E I \pdiff{^4 u}{z^4} + T_{DC} \pdiff{^2 u}{z^2} + K_{elec}^{dc} + F_{mag}\delta(z - z_0) = 0 \label{eqS:DCequation}\\
- E I \pdiff{^4 v}{z^4} + T_{DC} \pdiff{^2 v}{z^2} + T_{AC} \pdiff{^2 u}{z^2} + K_{elec}^{ac} e^{i \omega t} - \eta \pdiff{v}{t} = \rho A \pdiff{^2 v}{t^2} \label{eqS:ACequation}
\end{align}

To simplify computations, scale the parameters
\begin{gather*}
z' = \frac{z}{L}, u' = \frac{u}{r}, v' = \frac{v}{r}, T'_{dc} = \frac{L^2 T_{dc}}{E I}, k' = \sqrt{T'_{dc}}, f'_{dc} = \frac{K_{elec}^{dc} L^4}{r E I}, f'_{ac} = \frac{K_{elec}^{ac} L^4}{r E I} \\ f'_{mag} = \frac{F_{mag} L^3}{r E I}, \lambda = \frac{1}{L^2} \sqrt{\frac{E I}{\rho A}}, \eta' = \frac{\eta L^4}{\lambda E I}, \omega' = \frac{\omega}{\lambda}, t' = t \lambda
\end{gather*}

The scaled DC and AC governing equations can then be written as
\begin{align}
\pdiff{^4 u'}{{z'}^4} - T'_{dc}\pdiff{^2u'}{{z'}^2} &= f'_{dc} + f'_{mag}\delta(z'-z'_0) \label{eqS:scaledGovernDC} \\
\pdiff{^2 v'}{{t'}^2} - \eta'\pdiff{v'}{t'} + \pdiff{^4 v'}{{z'}^4} - T'_{dc} \pdiff{^2 v'}{{z'}^2} - T'_{ac}\pdiff{^2 u'}{{z'}^2} &= f'_{ac} e^{i \omega' t'} \label{eqS:scaledGovernAC}
\end{align}

To solve Equation (\ref{eqS:scaledGovernDC}), $T'_{dc}$ is taken as a constant, and doubly clamped boundary conditions are applied ($u'(0)=\pdiff{u'}{z'}(0)=u'(1)=\pdiff{u'}{z'}(1)=0$). The solution is then
\begin{multline} \label{eqS:uz}
u'(z') = \frac{f'_{dc}}{2{k'}^2}\left(\frac{\sinh(k')}{k'(\cosh(k')-1)}(\cosh(k'z')-1) - \frac{\sinh(k'z')}{k'} - ({z'}^2 - z')\right) + \\ \frac{f'_{mag}}{{k'}^3 \sigma_3} \left(\sigma_1 \left( \sinh(k'z')-k'z' \right) + \sigma_2 \left( \cosh(k'z') - 1 \right) \right) + \\ \frac{f'_{mag}}{{k'}^3}\left( \sinh(k'(z'-z'_0)) - k'(z'-z'_0) \right) \text{H}(z'-z'_0)
\end{multline}
where $\text{H}(z')$ is the Heaviside step function, and
\begin{align*}
\sigma_1 &= \cosh(k') - \cosh(k'z'_0) + \cosh(k'(1-z'_0)) - k'(1-z'_0)\sinh(k') - 1 \\
\sigma_2 &= \sinh(k'z'_0) - \sinh(k') + \sinh(k'(1-z'_0)) + k'z'_0 + k'(1-z'_0)\cosh(k') - k'\cosh(k'(1-z'_0))\\
\sigma_3 &= k'\sinh(k') - 2\cosh(k') + 2
\end{align*}
Equation (\ref{eqS:uz}) is used in Equation (\ref{eqS:T_dc}) to solve for a new value of $T'_{dc}$. The new $T'_{dc}$ is then substituted into Equation (\ref{eqS:uz}) again. This process is iterated until the $T'_{dc}$ result stabilizes.

To analyze the resonant motion of (\ref{eqS:scaledGovernAC}), consider that general motion can be decomposed into eigenstates of Equation (\ref{eqS:scaledGovernAC}),
\begin{equation} \label{eqS:motionDecomp}
v'(z',t') = \sum_n v'_n \xi'_n(z) e^{i \omega'_n t'}
\end{equation}
where $\xi'_n(z)$ is the amplitude profile of the $n^{th}$ eigenmode, which oscillates with frequency $\omega'_n$. The decomposition can be inserted into Equation (\ref{eqS:scaledGovernAC}) to determine equations for $\xi'_n(z)$ and $\omega'_n$. In general, $T'_{ac}$ will couple $\xi'_m$ to $\xi'_n$ for $m \neq n$, making solving difficult. However, for high Q resonators and near resonance excitation, $v'_m$ is very small for all $m \neq n$. Therefore, the mode coupling caused by $T'_{ac}$ can be neglected, and we get
\begin{equation} \label{eqS:xiGovern}
v'_n \left[ \left(-{\omega'}^2 + i \eta' \omega'\right)\xi'_n(z') + \pdiff{^4 \xi'_n}{{z'}^4} - T'_{dc}\pdiff{^2 \xi'_n}{{z'}^2} - {T'}^n_{ac} \pdiff{^2 u'}{{z'}^2} \right] = f'_{ac}
\end{equation}
where ${T'}_{ac}^n = 4 \int_0^1 \pdiff{u'}{z'}\pdiff{\xi'_n}{z'} dz'$ is the amplitude of the oscillatory $T'_{ac}$ for mode $\xi'_n$.

Noting that ${T'}_{ac}^n$ is independent of $v'_n$, it is treated as a constant, in which case (\ref{eqS:xiGovern}) describes the amplitude response of a damped driven harmonic oscillator if we can write
\begin{equation} \label{eqS:omegaMaster}
\pdiff{^4 \xi'_n}{{z'}^4} - T'_{dc}\pdiff{^2 \xi'_n}{{z'}^2} - {T'}_{ac}^n \pdiff{^2 u}{{z'}^2} = {\omega'}_n^2 \xi'_n
\end{equation}

To identify resonant frequencies of the CNT resonator, we must find $\omega'_n$ which have solutions in Equation (\ref{eqS:omegaMaster}). The resonant mode shape $\xi'_n$ must be solved piecewise, due to discontinuities introduced by the point force. The mode shape will have homogeneous components (with ${T'}^n_{ac} = 0$) and a particular solution for finite ${T'}^n_{ac}$.

Away from the discontinuity at $z' = z'_0$, the homogeneous solution is of the form
\begin{equation}
{\xi'}^h_n = A_1 \cos(k'_+ z') + A_2 \sin(k'_+ z') + A_3 \cosh(k'_- z') + A_4 \sinh(k'_- z')
\end{equation}
where $k'_\pm = \frac{1}{\sqrt{2}} \sqrt{\sqrt{{T'}_{dc}^2 + 4{\omega'_n}^2} \mp T'_{dc}}$.
The particular solution is
\begin{equation}
{\xi'}^p_n = A_5 \pdiff{^2 u'}{{z'}^2}
\end{equation}

To simplify computation for $z' > z'_0$, we substitute  $\left( 1 - z' \right)$ for $z'$ in the homogeneous solution. Thus, the full mode shape solution is
\begin{equation} \label{eqS:xi}
\xi'_n = \begin{cases} \begin{aligned} A_1 \cos(k'_+ z') + A_2 \sin(k'_+ z') + A_3 \cosh(k'_- z') + A_4 \sinh(k'_- z') + A_5 \pdiff{^2 u'}{{z'}^2} \end{aligned} & \text{if } z' \leq z'_0 \\ \begin{aligned}
B_1 \cos(k'_+ \left(1-z'\right)) + &B_2 \sin(k'_+ \left(1-z'\right)) + B_3 \cosh(k'_- \left(1-z'\right))\\ &\qquad\qquad + B_4 \sinh(k'_- \left(1-z'\right)) + B_5 \pdiff{^2 u'}{{z'}^2} \end{aligned} & \text{if } z' > z'_0 \end{cases}
\end{equation}

To determine the coefficients, we make use of the doubly clamped boundary conditions, matching conditions at $z' = z'_0$ and Equation (\ref{eqS:omegaMaster}). The boundary conditions are
\begin{equation*}
\xi'_n(0) = 0, \left.\pdiff{\xi'_n}{z'}\right\vert_{z'=0} = 0, \xi'_n(1) = 0, \left.\pdiff{\xi'_n}{z'}\right\vert_{z'=1} = 0
\end{equation*}
As the only force in Equation (\ref{eqS:scaledGovernAC}) is the uniform electric force, the AC modeshape and its derivatives will be continuous up to and including the third derivative, providing four matching conditions. Finally, substituting Equation (\ref{eqS:xi}) into Equation (\ref{eqS:omegaMaster}), gives an independent equation for both $z' < z'_0$ and $z' > z'_0$,
\begin{align*}
{T'}^n_{ac} + \omega_n^2 A_5 = 0 \\
{T'}^n_{ac} + \omega_n^2 B_5 = 0 \\
\implies A_5 = B_5
\end{align*}
Thus the mode shape coefficients are given by the  solutions to

\begin{equation} \label{eqS:systemEquationsOmega}
\begin{split}
\left[ \begin{array}{ccccccccc}
1 & 0 & 1 & 0 & 0 & 0 & 0 & 0 &  \left. \pdiff{^2 u'}{{z'}^2}\right\vert_{z'=0} \\
0 & k'_+ & 0 & k'_- & 0 & 0 & 0 & 0 &  \left. \pdiff{^3 u'}{{z'}^3}\right\vert_{z'=0} \\
0 & 0 & 0 & 0 & 1 & 0 & 1 & 0 &  \left. \pdiff{^2 u'}{{z'}^2}\right\vert_{z'=1} \\
0 & 0 & 0 & 0 & 0 & -k'_+ & 0 & -k_- &  \left. \pdiff{^3 u'}{{z'}^3}\right\vert_{z'=1} \\
c & s & ch & sh & -c_- & -s_- & -ch_- & -sh_- & 0 \\
-k'_+ s & k'_+ c & k'_- {sh} & k'_- ch & -k'_+ s_- & k'_+ c_- & k'_- sh_- & k'_- ch_- &  \Delta_3 \\
-{k'}_+^2 c & -{k'}_+^2 s & {k'}_-^2 ch & {k'}_-^2 sh & {k'}_+^2 c_- & {k'}_+^2 s_- & - {k'}_-^2 ch_- & -{k'}_-^2 sh_- & 0 \\
{k'}_+^3 s & -{k'}_+^3 c & {k'}_-^3 {sh} & {k'}_-^3 ch & {k'}_+^3 s_- & -{k'}_+^3 c_- & {k'}_-^3 sh_- & {k'}_-^3 ch_- &  \Delta_5 \\
{T'}_{ac}^{A_1} & {T'}_{ac}^{A_2} & {T'}_{ac}^{A_3} & {T'}_{ac}^{A_4} & {T'}_{ac}^{B_1} & {T'}_{ac}^{B_2} & {T'}_{ac}^{B_3} & {T'}_{ac}^{B_4} & {T'}_{ac}^{A_5} + {\omega'}_n^2
\end{array} \right]
\left[ \begin{array}{c}
A_1 \\ A_2 \\ A_3 \\ A_4 \\ B_1 \\ B_2 \\ B_3 \\ B_4 \\ A_5
\end{array}\right]
=
\left[ \begin{array}{c}
0 \\ 0 \\ 0 \\ 0 \\ 0 \\ 0 \\ 0 \\ 0 \\ 0
\end{array}\right]\end{split}
\end{equation}

where 
\begin{gather*}
c \equiv \cos(k'_+ z'_0), s \equiv \sin(k'_+ z'_0), ch \equiv \cosh(k'_- z'_0), sh \equiv \sinh(k'_- z'_0) \\ c_- \equiv \cos(k'_+ (1 - z'_0)), s_- \equiv \sin(k'_+ (1 - z'_0)), ch_- \equiv \cosh(k'_- (1 - z'_0)), sh_- \equiv \sinh(k'_- (1 - z'_0))
\end{gather*}
\begin{align*}
\Delta_3 &= \left. \pdiff{^3 u'}{{z'}^3}\right\vert_{z'=a^-} - \left. \pdiff{^3 u'}{{z'}^3}\right\vert_{z'=a^+} \\
\Delta_5 &= \left. \pdiff{^5 u'}{{z'}^5}\right\vert_{z'=a^-} - \left. \pdiff{^5 u'}{{z'}^5}\right\vert_{z'=a^+}
\end{align*}
and ${T'}_{ac}^X$ is the component of ${T'}^n_{ac}$ with coefficient $X$,
\begin{align*}
{T'}^{A_1}_{ac} &= - 4 k'_+ \int_0^{z'_0} \pdiff{u'}{z'} \sin(k'_+ z') dz' \\
{T'}^{A_2}_{ac} &= 4 k'_+ \int_0^{z'_0} \pdiff{u'}{z'} \cos(k'_+ z') dz' \\
{T'}^{A_3}_{ac} &= 4 k'_- \int_0^{z'_0} \pdiff{u'}{z'} \sinh(k'_+ z') dz' \\
{T'}^{A_4}_{ac} &= 4 k'_- \int_0^{z'_0} \pdiff{u'}{z'} \cosh(k'_+ z') dz' \\
{T'}^{B_1}_{ac} &= 4 k'_+ \int_{z'_0}^1 \pdiff{u'}{z'} \sin(k'_+ (1 - z')) dz' \\
{T'}^{B_2}_{ac} &= -4 k'_+ \int_{z'_0}^1 \pdiff{u'}{z'} \cos(k'_+ (1 - z')) dz' \\
{T'}^{B_3}_{ac} &= -4 k'_- \int_{z'_0}^1 \pdiff{u'}{z'} \sinh(k'_+ (1 - z')) dz' \\
{T'}^{B_4}_{ac} &= -4 k'_- \int_{z'_0}^1 \pdiff{u'}{z'} \cosh(k'_+ (1 - z')) dz' \\
{T'}^{A_5}_{ac} &= 4 \int_0^1 \pdiff{u'}{z'} \pdiff{^3 u'}{{z'}^3} dz'
\end{align*}

The resonant frequencies, $\omega'_n$, are those values which allow non-zero solutions to Equation (\ref{eqS:systemEquationsOmega}). Thus, to determine the resonant frequencies of the CNT resonator, we numerically solve for $\omega'_n$ which cause the 9x9 coefficient matrix to have zero determinant.

\subsection*{Parameter dependence of the maximum frequency shift}
To determine the relationship between device parameters and the maximum frequency shift, we determined the maximum shift for a range of device values and used numerical fitting to determine the corresponding relationship. The calculations below used initial parameters of $L = 1\:\mu\text{m}$, $r = 0.5\:\text{nm}$, $z_0 = \frac{L}{2}$,$T_0 = 0$, $h = 200\:\text{nm}$, and a magnetic field gradient of $3\cdot 10^5\, \text{T/m}$. All of the fittings shown below have $R^2 = 1.000$ with respect to the displayed data points.

The length dependence is shown in Figure \ref{fig:Lfit}, using $L$ from 600 nm to 2 $\mu$m. The calculated fitting is
\begin{equation}
\max(\Delta f) \propto L^{1.000}
\end{equation}

Next, diameter dependence is examined with $d$ ranging from 1 nm to 2.5 nm. The maximum frequency shift relationship is shown in Figure \ref{fig:dfit}. The calculated fitting is
\begin{equation}
\max(\Delta f) \propto d^{-3.500}
\end{equation}

The maximum shift for residual tension ranging from $-35 L^2/ (E I)$ to $35 L^2 / (E I)$ is shown in Figure \ref{fig:T0fit}. The calculated fitting is
\begin{equation}
\max(\Delta f) \propto \left(T'_0 + 4\pi^2\right)^{-1.027}
\end{equation}
where $T'_0 = T_0 L^2 / (E I)$.

\begin{figure}[h]
\centering
\includegraphics[scale = 0.88]{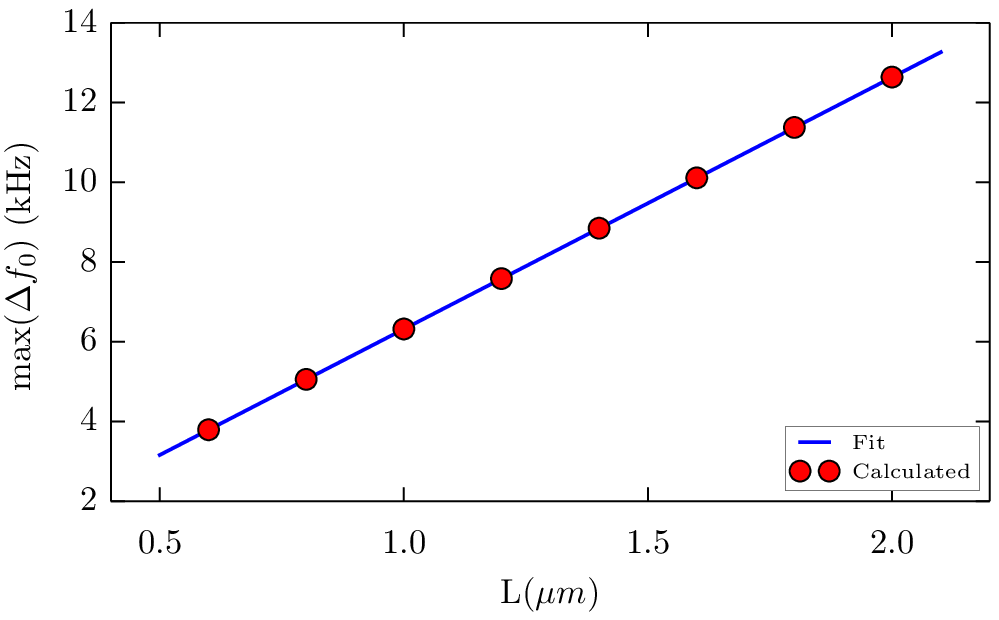}
\caption[FIXME]{Calculated maximum frequency shift as a function of resonator length, $L$, and the calculated fit.}
\label{fig:Lfit}
\end{figure}

\begin{figure}[h]
\centering
\includegraphics[scale = 0.88]{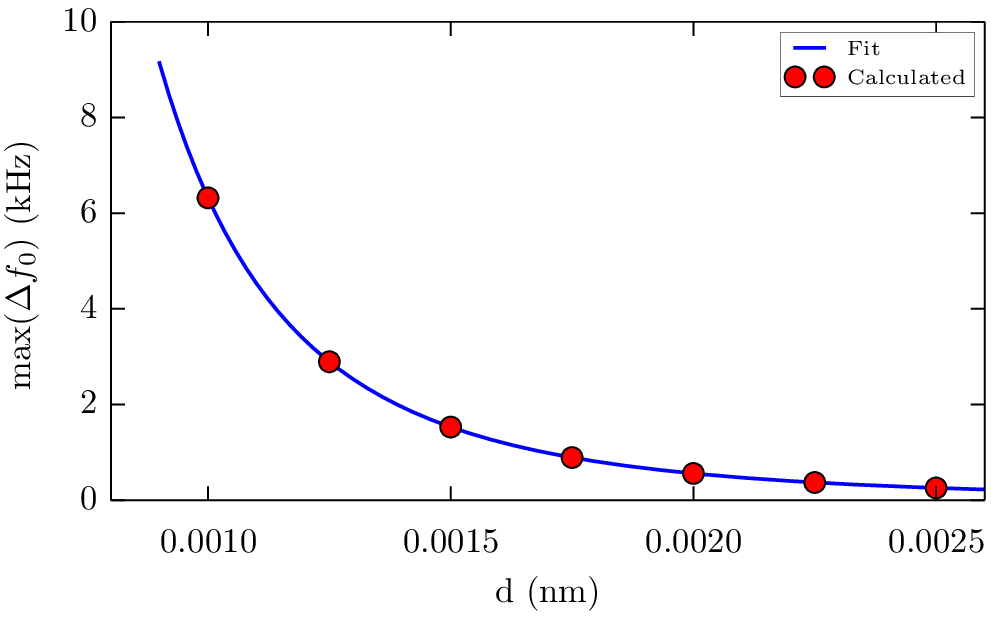}
\caption[FIXME]{Calculated maximum frequency shift as a function of CNT diameter, $d$, and the calculated fit.}
\label{fig:dfit}
\end{figure}

\begin{figure}[h]
\centering
\includegraphics[scale = 0.88]{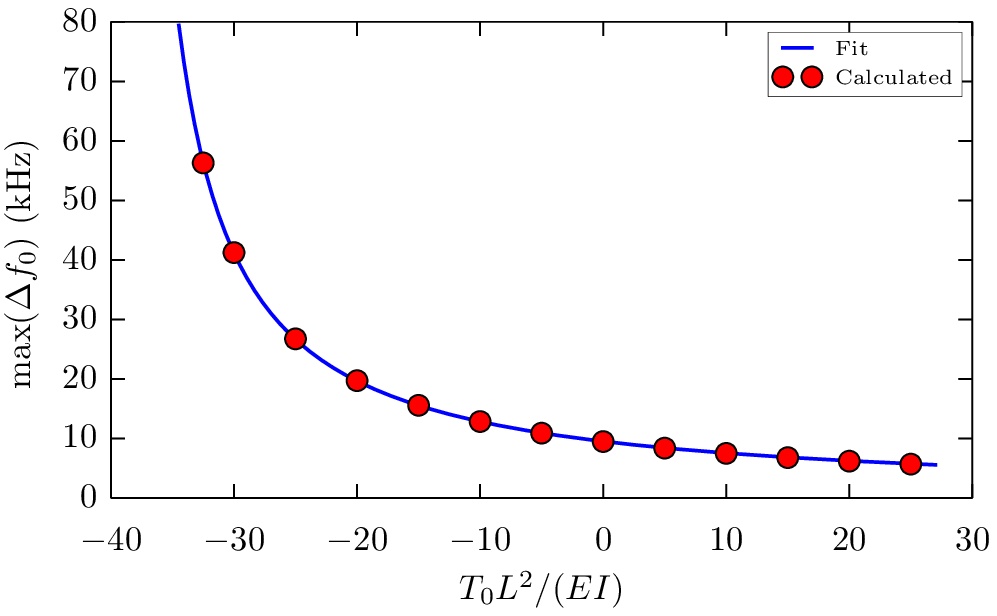}
\caption[FIXME]{Calculated maximum frequency shift as a function of residual tension, $T_0$, and the calculated fit.}
\label{fig:T0fit}
\end{figure}
\pagebreak
\end{widetext}

\end{document}